\documentclass[12pt]{article}
\usepackage{amsmath,amssymb,amsfonts,epsfig,graphicx,euscript}%
\usepackage{color}

\newcommand{\bse}{\begin{subequations}}
\newcommand{\ese}{\end{subequations}}
\newcommand{\be}{\begin{equation}}
\newcommand{\ee}{\end{equation}}
\newcommand{\bea}{\begin{eqnarray}}
\newcommand{\eea}{\end{eqnarray}}
\newcommand{\ba}{\begin{array}}
\newcommand{\ea}{\end{array}}
\newcommand{\h}{\frac{1}{2}}

\begin{document}
\hfill%
\vspace{1cm}
\begin{center}
{ \Large{\textbf{Non-equilibrium Phase Transition \\
from AdS/CFT}}} 
\vspace*{2cm}
\begin{center}
{\bf Mohammad Ali-Akbari\footnote{aliakbari@theory.ipm.ac.ir}, Ali Vahedi $^{1,}$\footnote{vahedi@ipm.ir}}\\%
\vspace*{0.4cm}
{\it {${}^1$School of Particles and Accelerators,\\ Institute for Research in Fundamental Sciences (IPM),\\
P.O.Box 19395-5531, Tehran, Iran}}  \\
{\it {${}^2$ Department of physics, Kharazmi university, \\
P.O.Box 31979-37551, Tehran, Iran}}  \\

\vspace*{1.5cm}
\end{center}
\end{center}

\vspace{.5cm}
\bigskip
\begin{center}
\textbf{Abstract}
\end{center}
Using AdS/CFT correspondence we study a non-equilibrium phase
transition in the presence of a constant external magnetic field.
The transition occurs when the sign of the differential conductivity
reverses. Utilizing numerical method we show that the type of the
transition depends on the value of magnetic field as well as the
temperature of the gauge theory. Moreover we show that this
transition does not depend on supersymmetry preserved by the system
at zero temperature and the dimension of the subspace on which the
fundamental matter field lives.

\newpage

\tableofcontents

\section{Introduction}
One of the most challenging open problems in theoretical physics is
to study a physical system at non-equilibrium phase. In fact when a system
is in equilibrium one may describe it without explicitly knowing
about the microscope details of the system. However, in the case of
non-equilibrium systems more information about dynamics is needed in order to understand
the most basic properties of the system.  Nevertheless there are certain
cases where even though the systems are out of equilibrium, one
still has a good control over them. The first one is the standard
theory of non-equilibrium thermodynamics which is based on the local
equilibrium hypothesis that roughly states that each small part of
the non-equilibrium system can be considered as a system in equilibrium \cite{Balescu}.
Another family, which is extensively studied
in the literature, is called non-equilibrium steady states. These
systems are clearly out of equilibrium but they have no
macroscopically observable time dependence \cite{steady0, steady}. A
few example of such non-equilibrium systems were discussed in
\cite{steady1, Henkel}. In this paper, using AdS/CFT correspondence,
we consider a strongly coupled non-equilibrium steady state and we
are interested in studying a phase transition in the presence of
external magnetic field.

The AdS/CFT correspondence states that type IIB string theory on
$AdS_5\times S^5$ background, which describes near horizon geometry
of a stack of $N_c$ of extremal D3-branes, is dual to the
four-dimensional super-conformal Yang-Milles (SYM) theory with gauge
group $SU(N_c)$ \cite{ads/cft}. The correspondence is more applicable
in the limit of large $N_c$ and large 'tHooft coupling,
$\lambda=g_{YM}^2N_c$ where $g_{YM}$ is the gauge theory coupling
constant. In these limits it is well-known that a strongly coupled
SYM is dual to the IIb supergravity which is the low energy
effective theory of superstring theory. This correspondence has also been
generalized to the thermal SYM theories. As a result a strongly
coupled thermal SYM theory corresponds to the supergravity in an
AdS-Schwarzschild black hole background where SYM theory temperature
is identified with the Hawking temperature of the AdS black hole
\cite{Witten}.

Using the AdS/CFT idea, a \textit{non-linear} conductivity has been
computed in \cite{Karch, O'Bannon}\footnote{Similar non-equilibrium
systems have been studied in \cite{Alam:2012fw}. }. Matters(or
quarks) in the fundamental representation of the $SU(N_c)$ gauge
group must be added to the original SYM theory to compute the
conductivity. On the gravity side it can be done by introducing
$N_f$ flavor D$p$-branes (in this paper we consider $p=5,7$) in the
probe limit \cite{Karch:2002sh}. The probe limit means that the
number of flavor D$p$-branes are much smaller than the number of
D3-branes. According to the AdS/CFT dictionary, the local $U(N_f)$
symmetry on the flavor D7-branes corresponds to global $U(N_f)$
symmetry whose $U(1)_B$ subgroup is identified with baryon number.
Then the fundamental matters are couples to external electric and
magnetic fields and therefore we expect a non-zero current, $J$.
Then the conductivity is defined as $\sigma=\frac{J}{E}$ where $E$
is an external electric field.

An interesting property of this non-linear conductivity has been
studied in \cite{Nakamura1, Nakamura2}. In fact, due to the
non-linearity of conductivity, differential conductivity
$\frac{\partial J}{\partial E}$ may be either positive or negative.
It was shown that both signs of the differential conductivity happen
in the D3-D7 system depending on the temperature of the system. Moreover,
it was discussed that this sign conversion can be considered as a
transition or a continuous crossover. At the critical temperature
$T_c$, a second order phase transition from negative differential
conductivity (NDC) to positive differential conductivity (PDC)
occurs. Above the critical temperature, the system represents a
first order phase transition. When $T<T_c$, there is a continuous
crossover by which we mean that there is no sharp transition from
NDC to PDC (or vise versa).

In this paper the main question of interest is how the transition
between NDC and PDC can be affected by an external magnetic field.
For this reason in next section we review the non-linear
conductivity which depends on the constant magnetic field and study
the various types of transition for D3-D7 system in section 3. Then
we consider two D3-D5 systems. The subspace on which the fundamental
matter lives is a main difference between these two D3-D5 systems.
Moreover, the supersymmetry does not preserve by one of them even
\textit{at zero temperature}.

\section{Review on non-linear conductivity}
In order to compute the conductivity, let us start with D3-D7 system
which is a supersymmetric intersection of $N_c$ D3-branes and $N_f$
flavor
D7-branes as 
\be %
\begin{array}{ccccccccccc}
                   & t & x & y & w & z &  S^3 & \theta & \psi  \\
                  D3 & \times & \times & \times & \times &  &  &  & \\
                  D7 & \times & \times & \times & \times & \times & \times &  &
\end{array}
\ee %
Here we consider probe limit which means that $N_f\ll N_c$. In other
words, the back-reaction of the probe D7-branes is ignored in this limit
and consequently the dynamics of D3-D7 system is given by the
dynamics of the probe branes. This configuration holographicly describes
a SYM theory with the gauge group $SU(N_c)$ coupled to the matter
field in the fundamental representation of the gauge group. The
mass of the matter field is considered as a distance between D3-
and D7-branes in the transverse plane. If this distance is zero, the
matter field is massless.

In the large $N_c$ and large t'Hooft coupling limits,
$\lambda=g_{YM}^2N_c=4\pi g_s N_c$, where $g_s$ is string coupling
constant, the D3-branes may be replaced by $AdS_5\times S^5$
background which is dual to a strongly coupled SYM theory at zero
temperature. On the other hand at finite temperature the $AdS_5$ is
replaced by AdS-Schwarzschild black hole where the Hawking
temperature of AdS black hole is identified with the temperature of
the strongly coupled SYM theory. The AdS-Schwarzschild metric
in units of the $AdS_5$ radius is%
\be\label{background} %
 ds^2=\frac{dz^2}{z^2}-\frac{1}{z^2}\frac{(1-z^4/z_h^4)^2}{1+z^4/z_h^4}dt^2+\frac{1}{z^2}\left(1+\frac{z^4}{z_h^4}\right)d\vec{x}^2+d\Omega_5^2,
\ee %
and the metric of $S^5$ is %
\be\label{background1} %
 d\Omega_5^2=d\theta^2+\sin^2\theta d\psi^2+\cos^2\theta d\Omega_3^2, %
\ee %
where $z$ and $t$ are the radial and time directions, respectively.
$x, y$ and $w$ denote a three dimensional Euclidean space. In this
coordinate boundary is located at $z=0$. The horizon is at $z_h$ and
the Hawking temperature of black hole is then given by
$T=\frac{\sqrt{2}}{\pi z_h}$. There is also a form field %
\be %
g_s C^{(4)} = u^4 dt\wedge dx \wedge dy \wedge dw, %
\ee %
which couples to the D3-branes.

The dynamics of $N_f$ D7-branes in an arbitrary
background is described by Dirac-Born-Infeld(DBI) and Chern-Simons(CS) actions %
\be\begin{split}\label{action} %
 S &= S_{\rm{DBI}}+ S_{\rm{CS}}~,\cr
 S_{{\rm{DBI}}}&=-N_fT_{D7}\int d^8\xi\
 e^{-\phi}\sqrt{-\det(g_{ab}+B_{ab}+2\pi\alpha'F_{ab})}~,\cr
 S_{\rm{CS}} &=N_fT_{D7}\int P[\Sigma C^{(n)}e^B]e^{2\pi\alpha'F}~,
\end{split}\ee %
where the induced metric, $g_{ab}$, and Kalb-Ramond field, $B_{ab}$, are given by %
\be\begin{split} %
 g_{ab}=G_{MN}\partial_a X^M\partial_b X^N,\cr
 B_{ab}=B_{MN}\partial_a X^M\partial_b X^N.
\end{split}\ee %
$T_{D7}^{-1}=(2\pi)^7l_s^8g_s$ is the D7-brane tension. $\xi^a$ are
worldvolume coordinates and the capital indices $M,N,...$ are used
to denote space-time coordinates. In our case the background metric
$G_{MN}$ is given in \eqref{background}. $F_{ab}$ is the field
strength of gauge fields living on the D7-branes. We use static
gauge which means that the D7-branes are extended along
$t,x,y,w,z,S^3$ or equivalently along $AdS_5\times S^3$. In the CS
action, $C^{(n)}$ denotes Ramond-Ramond form fields and $P[...]$ is
the pull-back of the bulk fields to the worldvolume of D7-branes.

In order to describe a nonlinear conductivity we consider
appropriate field configurations on the D7-branes as follows
\footnote{A more general case in which $A_t(z)\neq0$ and
$A_y(x,z)=Bx+a_y(z)$ was considered in \cite{O'Bannon}. According to
the results of \cite{Nakamura1}, pair-creation process is essential
for NDC. In order to study the phase transition between NDC and PDC,
we safely chose $A_t(z)=0$.
Just for simplicity, we also set $A_y(x)=Bx$.} %
\be\label{configuration} %
 A_x(t,z)=-E t+a_x(z),\ A_y(x)= B x,\ \theta(z).%
\ee %
Substituting the field configurations \eqref{configuration} into the
action \eqref{action}, and performing two Legendre transformations
one arrives at (for more details see \cite{O'Bannon})
\be\begin{split} %
 S_{D7}&=\int dtdz\  {\cal{L}}\cr
 &=-\frac{1}{2\pi\alpha'}\int dtdz\
 g_{zz}^{1/2}|g_{tt}|^{-1/2}g_{xx}^{-1}\sqrt{\xi\chi},
\end{split}\ee %
where %
\bse\begin{align}%
 \xi&=|g_{tt}|g_{xx}^2+(2\pi\alpha')^2(|g_{tt}|B^2-g_{xx}E^2),\\
 \chi&={\cal{N}}^2(2\pi\alpha')^2|g_{tt}|g_{xx}^2\cos^6\theta-C^2,
\end{align}\ese %
and ${\cal{N}}=\frac{\lambda N_c N_f}{(2\pi)^4}$. Since both sides
of the above action are divided by the volume of $\mathbb{R}^3$ so
this is indeed an action density. Note that $C=\frac{\partial
L}{\partial(\partial_z A_x)}$ is the conserved charge associated with
$A_x$. Notice that the CS action does not contribute to the action
of D7-branes.

The reality condition of action imposes that the two
functions $\xi$ and $\chi$ must
vanish at the same point, say, at $z=z_*$. Therefore, we have %
\bse\label{coductivity}\begin{align}%
 \label{coductivity1}|g_{tt}|g_{xx}^2+(2\pi\alpha')^2(|g_{tt}|B^2-g_{xx}E^2)&=0,\\
 \label{coductivity2}{\cal{N}}^2(2\pi\alpha')^2|g_{tt}|g_{xx}^2\cos^6\theta(z_*)-C^2&=0,
\end{align}\ese %
where all functions of $z$ are evaluated at $z_*$.

The field configurations \eqref{configuration} can be expanded near
the boundary to extract the field theory information. They
asymptotically approach
the boundary as \cite{O'Bannon}%
\be\begin{split} %
 A_x(t,z)&=-E t + \h\frac{\langle
 J\rangle}{{\cal{N}}(2\pi\alpha')^2}z^2+O(z^4),\cr
 \theta(z)&=\theta_0 z+\theta_2 z^3+O(z^5).
\end{split}\ee%
According to the AdS/CFT correspondence, for different fields
non-normalizable and normalizable modes in the asymptotic expansion
near boundary define the source for the dual operator and its
expectation value, respectively. Therefore, for $A_x$ the
non-normalizable term is proportional to electric field $E=F_{tx}$ and
$A_y$ induces a magnetic field $B=F_{xy}$ on the boundary. The
$A_x$'s normalizable term is proportional to the expectation value
of corresponding dual operator,  $\langle J\rangle$, in the gauge
theory. In \cite{Karch, O'Bannon} it was shown that $C=\langle
J\rangle$. The non-normalizable term of $\theta(z)$ corresponds to the
mass of fundamental matters, $(2\pi\alpha')m=\theta_0$, and the
normalizable term is proportional to $\langle O_m\rangle$ where
$O_m$ is the dual operator to mass.

Using \eqref{coductivity}, one can now find the expectation value of
the current in the $x$ direction. \eqref{coductivity1} gives $z_*$
in terms of $z_h$, electric and magnetic fields as
\be\begin{split}\label{condtion} %
 z_*^4=\left(2F(e,b)-\sqrt{(2F(e,b)-1)^2-1}-1\right)z_h^4,
\end{split}\ee %
where %
\be\begin{split}
 F(e,b)&=\h\left(1+e^2-b^2+\sqrt{(e^2-b^2)^2+2(e^2+b^2)+1}\right),\cr %
 e&=\frac{E}{\frac{\pi}{2}\sqrt{\lambda}T},\ \
 b=\frac{B}{\frac{\pi}{2}\sqrt{\lambda}T}.
\end{split}\ee
Substituting $z_*$ into \eqref{coductivity2}, the expectation value of current becomes %
\be %
 \langle
 J\rangle^2={\cal{N}}^2(2\pi\alpha')^2|g_{tt}|g_{xx}^2\cos^6\theta(z_*).
\ee %
Then the conductivity is simply given by %
\be %
\sigma=\frac{\langle J\rangle}{E}.
\ee %
In the case of $B=0$, it was shown in \cite{Nakamura2} that in the
small-$J$ region the above system exhibits NDC. However, PDC is seen
in the large-$J$ region. In other words, there is a transition
between NDC and PDC. The question we would like to pose is how an
external magnetic field affects this transition.

\section{Non-equilibrium phase transition}
In this section we study the behavior of the current $J\equiv\langle
J\rangle$ in the D3-D7 and the D3-D5 systems when an external
magnetic field is turned on.

\subsection{D3-D7 system}
Since we need to solve the equation of motion for $\theta(z)$
numerically, let us explain our numerical method. The equation of
motion is a second-order differential equation and hence we need two
initial conditions to solve it. In order to specify these initial
conditions, for given $E$ and $B$ we compute $z_*$. Then we assign
$0<\theta(z_*)<\pi/2$ and compute $J$ (or vice versa). Therefore,
the first initial condition is $\theta(z_*)$. Using the equation of
motion for $\theta(z)$ the second condition, which is
$\theta'(z_*)$, can be found. We
expand $\theta(z)$ around $z_*$ as %
\be %
\theta(z)=\theta(z_*)+\theta'(z_*)(z-z_*)+\frac{1}{2}\theta''(z_*)(z-z_*)^2+...\
.
\ee %
The equation of motion at zero order automatically vanishes and
$\theta'(z_*)$ will be found from the first order term. However, the details
are cumbersome and unilluminating and therefore we will not present
it here. After finding the $\theta(z)$, its asymptotic value is
proportional to the mass of fundamental matters. Note that in all
subsequent plots we have chosen $\lambda=(2\pi)^2$ and $N_f N_c=40$.

In the presence of $AdS$-Schwarzschild black hole, when the electric
field is non-zero, the various embeddings of probe D7-branes can be
classified into three groups \cite{Kobayashi, Mateos, Kim:2011qh,
Kim:2011qh1}. The Minkowski embedding without horizon(ME) are those
embeddings where the probe D7-branes close off above the horizon and
$z_*$. In other words, the $S^3$ part of D7-branes shrinks to zero
at an arbitrary value $z_s<z_*<z_h$. Conversely if $z_*<z_s<z_h$,
the embedding is called Minkowski embedding with horizon(MEH). In
fact $z_*$ can be considered as a worldvolume horizon for MEHs
\cite{Kim:2011qh1}. As it can be seen from \eqref{condtion}, in the
zero electric field limit $z_*=z_h$ and therefore ME and MEH come to
be identical. The third group is black hole embedding(BE). In this
group the $S^3$ part of D7-branes shrinks but does not reach zero
size for $z_s\leqslant z_h$. On the ME the quark-antiquark bound
states are stable indicating that there is no charge carrier and
hence the system behaves as an insulator. On the contrary, on the BE
and MEH the bound states are unstable and as a result non-zero
conductivity is observed \cite{Kim:2011qh1, Erdmenger}.

\begin{figure}[ht]
\begin{center}
\includegraphics[width=2.6 in]{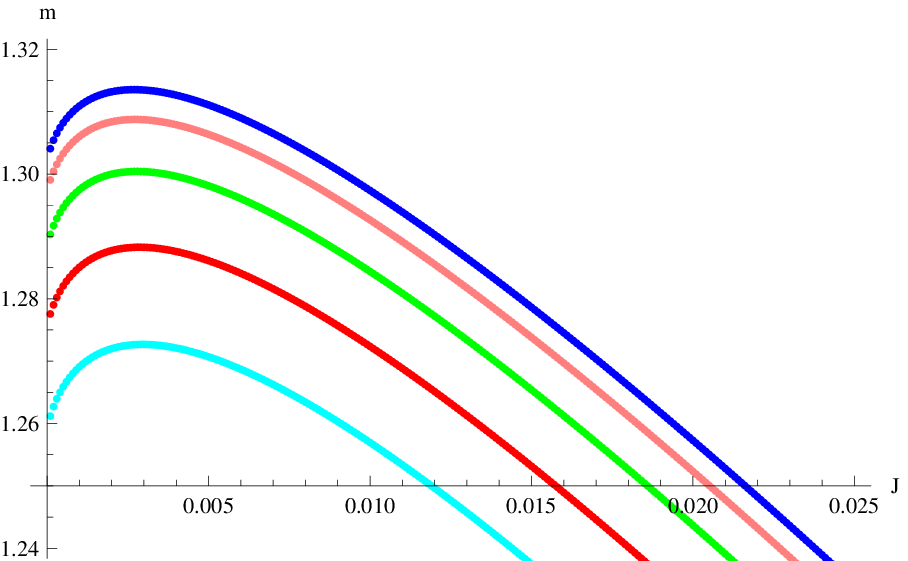}
\hspace{2mm}
\includegraphics[width=2.6 in]{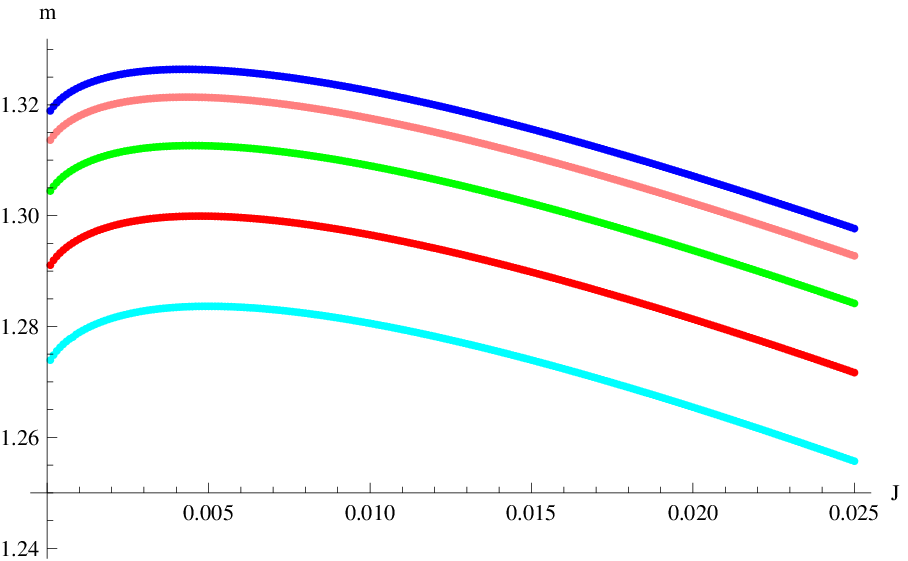}
\caption{The value of mass as a function of current has been plotted
for B=0, 0.3, 0.5, 0.7, 0.9 (top to bottom) and $T=0.450158$. We set $E=0.1$ ($E=0.2$) in
the left (right) figure.\label{E-B}}
\end{center}
\end{figure}%

As it was already mentioned, the mass of matter field corresponds to
the asymptotic value of $\theta(z)$. In the zero electric field
case, at a fixed temperature $T$, the ratio of the mass to the
temperature determines which embedding is thermodynamically favored.
It turns out that for sufficiently small $m/T$, BE is favorable
while ME will be chosen for sufficiently large $m/T$
\cite{CasalderreySolana}. Similarly in  our case in figure \ref{E-B}
for given $E$, $B$ and temperature it is easily seen that there is a
maximum value for the mass, $m_{max}$, indicating that ME is more
desirable for $m>m_{max}$.
Figure \ref{E-B} also shows that when the electric field and
temperature are fixed the maximum value of mass decreases as the
magnetic field increases\footnote{The effect of electric field on
this system has been discussed in \cite{Nakamura1, Nakamura2}.}.
This can be explained if we notice that the effective tension of
probe D7-branes increases in the presence of magnetic
field\footnote{Consider a D7-brane in the flat background. In the
presence of magnetic field, the action becomes
\be %
S_{DBI}=-T_7\sqrt{1+B^2}\int d^{8}\xi=-\tau_{eff} \int d^{8}\xi,
\ee %
where $B=F_{xy}$ and all other fields have been turned off. It is
clearly seen that $\tau_{eff}\geq T_7$.}. Therefore, they resist
more against the shape deformation. In other words, compared to the
case with $B=0$, $z_s$ decreases. As a result, in order to have a
non-zero current, the asymptotic distance between D3- and D7-branes,
or equivalently $m$, must decrease. Moreover, in this figure it is
clearly seen that for a given value of mass there are two embeddings
where they have different currents. These embeddings are observed
for a range of currents depending on the values of magnetic field
and mass. As we will discuss later on, this indicates a phase
transition in the system.

\begin{figure}[ht]
\begin{center}
\includegraphics[width=2.6 in]{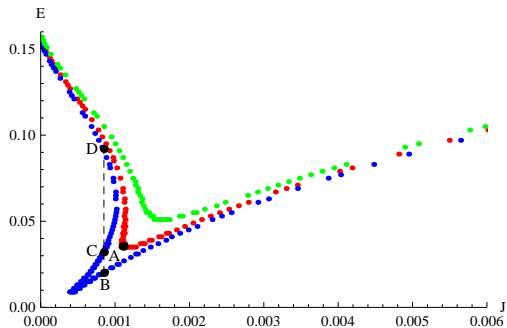}
\caption{For various temperatures the valve of electric field $E$
versus $J$ has been showed. The curves with $T=0.449350>T_c$,
$T_c=0.449250$ and $T=0.449100<T_c$ are presented by the blue, red and
green, respectively. We have chosen $m=1.302$ and $B=0.3$.
\label{phase}}
\end{center}
\end{figure}%

In figure \ref{phase} at a fixed magnetic field we have plotted the
value of the electric field versus current for different temperatures.
One can see that in a region with small current the differential
conductivity, defined by $\frac{\partial J}{\partial E}$, is
negative. But in a region with large current this quantity becomes
positive. This conversion corresponds to a transition or to a
continuous crossover between NDC and PDC. The green curve in which
$T<T_c$ shows that the region with NDC is smoothly connected to the
region with PDC. Therefore, a continuous \textit{crossover} rather
than a phase transition occurs. At the critical temperature,
$T=T_c$, the red curve shows that $\frac{\partial E}{\partial
J}\thicksim 0$ at $J=J_c\thicksim 0.0011$. Consequently, the
differential conductivity diverges at the critical point $A$. However,
$J/E$ is finite at the critical point and accordingly the conversion
represents a \textit{second-order} phase transition. The blue curve
shows that, in the region with small $J$, three values of the
electric field (points B, C and D in the figure \ref{phase}) are
allowed for a fixed current. It means that the electric field can
jump from one value to another and therefore NDC is discontinuously
connected to PDC. Since the electric field changes discontinuously,
it is acceptable to consider this transition as a \textit{first
order} phase transition.

In equilibrium the solutions with smaller free energy are the
preferred solutions. Since our system is not in equilibrium, we can
not define a free energy thermodynamically. Therefore, in our case,
a significant question is what quantity can be used to select the
favorable solution among the three solutions, for example among the
three points $B, C$ and $D$ in figure \ref{phase}. In
\cite{Nakamura2} it has been proposed that the energy of system may
be considered as a criteria to determine the favorable solution. In
the presence of
magnetic field the Hamiltonian density of system becomes%
\be\label{hamiltonian}\begin{split}
 {\cal{H}}&=\dot{A}_x\frac{\partial{\cal{L}}}{\partial\dot{A}_x}-{\cal{L}}\cr
 &=\sqrt{\frac{g_{zz}|g_{tt}|}{g_{xx}}
 \frac{({\cal{N}}^2(2\pi\alpha')^2|g_{tt}|g_{xx}^2\cos^6\theta(z)-J^2)(g_{xx}^2+(2\pi\alpha')^2B^2)^2}
 {|g_{tt}|g_{xx}^2-(2\pi\alpha')^2g_{xx}E^2+(2\pi\alpha')^2|g_{tt}|B^2}},
\end{split}\ee %
where $\dot{A}_x=\frac{\partial A_x}{\partial t}$.  Then the energy
of system is given by
\be %
E=\lim_{\epsilon\rightarrow0}\left(\int_{\epsilon}^{z_h}dz\
{\cal{H}}-H_c(\epsilon)\right).
\ee %
The suitable counter term for the Hamiltonian, $H_c$, is
equal to $L_1+L_2+L_f-H_F$ where\footnote{We would like to thank the referee for his/her comment
on the signature of the electric and magnetic fields in \eqref{120}.}
\be\label{120} %
H_F=\h {\cal{N}}(2\pi\alpha')^2(B^2+E^2)\log\epsilon + O(\epsilon^4
\log\epsilon),
\ee %
and the other terms have explicitly been given in \cite{O'Bannon}.
It numerically turns out that the solutions with the
\textit{largest} value of the electric field have the lowest energy.
For instance, among three points $B, C$ and $D$, the embedding
corresponding to point $D$ is energetically favorable\footnote{Our
numerical calculation is not reliable for $B<0.1$ and therefore this
result is applicable for $B\geq 0.1$.}.

\begin{figure}[ht]
\begin{center}
\includegraphics[width=2.6 in]{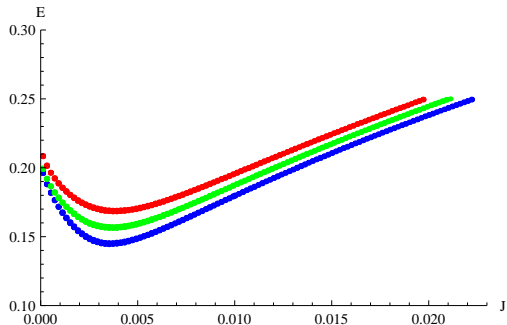}
\caption{The value of electric field as a function of current has been plotted
for B=0.2, 0.1, 0 (top to bottom), $T=0.449000$ and $m=1.315$. \label{E-J}}
\end{center}
\end{figure}%

In the crossover region between NDC and PDC, figure \ref{E-J}
suggests that a non-zero current is observed for $E>E_c\simeq0.14,
0.16, 0.17$ corresponding to $B=0, 0.1, 0.2$, respectively. More
precisely, for $E<E_c$, MEs are favorable and hence there is no
current. For $E\geq E_c$ the system is a conductor and then
$J\neq0$. This figure shows that by increasing the magnetic field,
the value of the $E_c$ also raises. The reason for this is as
follows. The MEs change to the BEs or to the MEHs in the presence of
sufficiently large electric field as it has been discussed in
\cite{Nakamura1,Nakamura2}. However, the magnetic field increases
the effective tension of probe D7-branes and therefore in order to
have a non-zero current we need a larger value of the electric
field. In fact electric and magnetic fields have opposite effect on
the tension of D7-branes. Notice that in this figure the mass has
been fixed.

\begin{figure}[ht]
\begin{center}
\includegraphics[width=2.6 in]{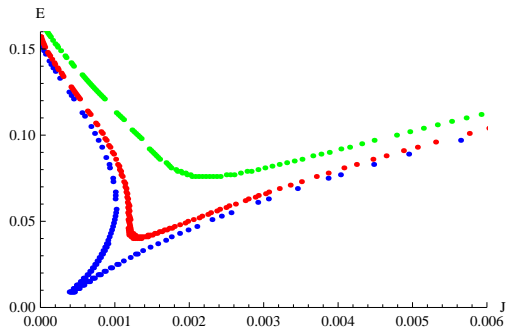}
\caption{At a fixed temperature, $T=0.449350$, the valve of electric
field $E$ versus $J$ has been showed. The different curves
correspond to different magnetic fields, with blue, red and green
corresponding to $B=0.30, 0.315$ and $0.36$, respectively. We set
$m=1.302$.   \label{phase2}}
\end{center}
\end{figure}%

\begin{figure}[ht]
\begin{center}
\includegraphics[width=2.6 in]{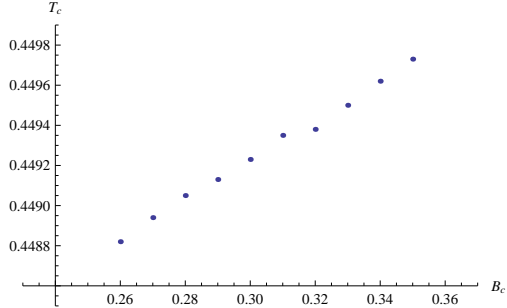}
\caption{$T_c$ as a function of $B_c$ for $m=1.302$. \label{Tc-Bc}}
\end{center}
\end{figure}

We showed that the type of transition between NDC and PDC can be a
phase transition (first or second) or a crossover depending on the
value of temperature. Now consider a first order phase transition
showing by the blue curve in figure \ref{phase2} for $B=0.30$. Note
that in this figure the mass and temperature are kept fixed. We
numerically observed that for larger values of the magnetic field,
the type of transition will alter. The transition is a second order
phase transition (crossover) when the magnitude of magnetic field is
$0.315$ ($0.36$). As a matter of fact, $B_c=0.315$ is a critical
value for the magnetic field presenting a first order phase
transition. For $B>B_c$ a crossover takes place and conversely if
$B<B_c$, we have a first order phase transition. As a result, the
type of transition depends not only on the temperature but also on
the value of external magnetic field. Figure \ref{Tc-Bc} shows that
$T_c$ depends almost linearly on $B_c$.

\subsection{D3-D5 System}
Here, we would like to investigate whether this transition depends
on the subspace that the fundamental matter lives and on the
supersymmetry which is preserved by the system at zero temperature.
In the first case, D5-branes are extended along a 2+1 dimensional
subspace. This system preserves supersymmetry at zero temperature.
In the other case, the subspace is 1+1 dimensional and the system is
not supersymmetric even at zero temperature. It is worthwhile to
notice that the supersymmetry is anyway broken at finite
temperature.

\subsubsection{D5 extended along a 2+1 dimensional subspace}
The supersymmetric system we are interested in is D3-D5 system which
can schematically be showed with the following intersection of $N_c$
D3-branes and $N_f$ flavor D5-branes
\be %
\begin{array}{ccccccccccc}
                   & t & x & y & w & z & S^2 & \theta & S^2   \\
                  D3 & \times & \times & \times & \times &  &  &  &  \\
                  D5 & \times & \times & \times &  & \times & \times  &  &
\end{array}
\ee %
\begin{figure}[ht]
\begin{center}
\includegraphics[width=2.6 in]{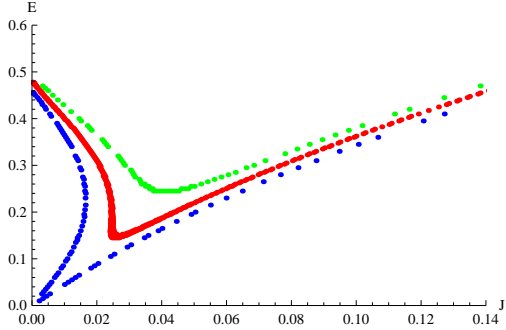}
\caption{The green ($T=0.446208$), the red ($T_c=0.447958$) and the
blue ($T=0.450158$) curves corresponding to a crossover, second and
first order phase transitions, respectively. We set $m=1.66$ and
$B=0.3$. \label{m-j-D5}}
\end{center}
\end{figure}%
As before, in the low energy limit, D3-branes are replaced by
$AdS_5\times S^5$ whose dual is SYM in $3+1$ dimensions with the
gauge group $SU(N_c)$. The D5-branes wrap an $AdS_4\times S^2$ and
give rise to matter fields in the fundamental representation. These
matter fields live on the intersection of the D5-branes and
D3-branes which is a $2+1$ dimensional subspace. However, degrees of
freedom of the SYM live in $3+1$ dimensions. In fact this is the
main difference between this system and the D3-D7 system. Similar to
the previous section the mass of matter fields is given by the
asymptotic value of $\theta(z)$.

In the probe limit, the dynamics of D5-branes is governed by DBI action, introduced
in \eqref{action}, in the $AdS_5\times S^5$ background. Compared to the action of
probe D7-branes, the only change is in $\chi$  \cite{O'Bannon}
\be
 \chi=(2\pi\alpha')^2g_{xx}\left({\cal{N}}^2 (2\pi\alpha')^2 |g_{tt}|g_{xx}\cos^4\theta(z)-
 J^2\right),
\ee %
where ${\cal{N}}=\frac{4\sqrt{\lambda}N_fN_c}{3(2\pi)^3}$.
Note that we parameterized the metric of the $S^5$ as %
\be %
 d\Omega_5^2=d\theta^2+\sin^2\theta d\Omega_2^2+\cos^2\theta d\Omega_2^2. %
\ee %
The reality condition of the action then leads to %
\be %
 J^2={\cal{N}}^2 (2\pi\alpha')^2 |g_{tt}|g_{xx}\cos^4\theta(z_*),
\ee %
and \eqref{condtion}.

We numerically observe that all results are qualitatively similar
to the case studied in the pervious section. In particular, we
observe that for a given value of mass two solutions with different
$J$s exist signaling a phase transition. A first (blue curve) and
second (red curve) order phase transitions and a continuous
crossover (green curve) have been plotted in figure \ref{m-j-D5}.

The Hamiltonian of system can be found by replacing
$g_{xx}\cos^6\theta(z_*)$ by $\cos^4\theta(z_*)$ in
\eqref{hamiltonian} and then the energy becomes
\be %
 E=\lim_{\epsilon\rightarrow0}\left(\int_{\epsilon}^{z_h}dz\
{\cal{H}}-H_c(\epsilon)\right),
\ee %
where $H_c$ can be found in \cite{Karch:2005ms}. We numerically
observe that the configuration with the larger value of electric
field is more favorable, energetically. Note that similar to the
D3-D7 case, there is a lower value for the magnetic field,
$B_{min}\simeq 0.05$, where our numerical calculation is reliable
for $B>B_{min}$.

\subsubsection{D5 extended along a 1+1 dimensional subspace}
Now consider the following system
\be %
\begin{array}{ccccccccccc}
                   & t & x & y & w & z & S^3 & \theta & \psi   \\
                  D3 & \times & \times & \times & \times &  &  &  &  \\
                  D5 & \times & \times &  &  & \times & \times  &  &
\end{array}
\ee %
\begin{figure}[ht]
\begin{center}
\includegraphics[width=2.6 in]{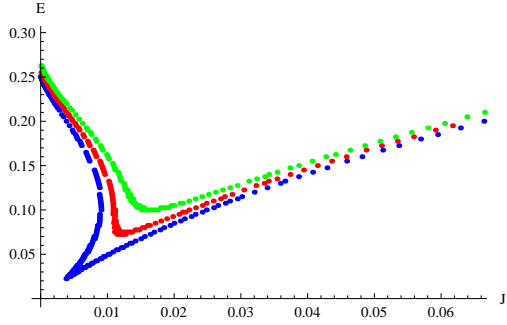}
\caption{The green ($T=0.446658$), the red ($T_c=0.447158$) and the
blue ($T=0.447658$) curves corresponding to a crossover, second and
first order phase transitions, respectively. We set $m=0.905$.
\label{phase-D5-D3-B0}}
\end{center}
\end{figure}%
In this case the matter fields are living in the 1+1 dimensional
subspace which is the intersection of D3- and D5-branes. The
$\chi$ then becomes %
\be %
\chi=(2\pi\alpha')^2g_{xx}\left({\cal{N}}^2 (2\pi\alpha')^2
|g_{tt}|\cos^6\theta(z)- J^2\right),
\ee %
where $\theta(z)$ has been introduced in \eqref{background1} and
${\cal{N}}=\frac{\sqrt{\lambda}N_fN_c}{(2\pi)^2}$. For this system
there is no magnetic field and therefore we consider an external
electric field and hence \eqref{condtion}
becomes%
\be %
z_*^2=\left(\sqrt{e^2+1}-e\right)z_h^2.
\ee %
Requiring the reality condition of the action leads to %
\be %
J^2={\cal{N}}^2 (2\pi\alpha')^2 |g_{tt}|\cos^6\theta(z_*).
\ee %
Our results, shown in the figure \ref{phase-D5-D3-B0}, show that the
transition between NDC and PDC occurs in this case, too.

\section{Conclusion}
In this paper we studied the effect of magnetic field on the non-equilibrium phase transition of the differential conductivity.
Our findings can be summarized as follows.
\begin{itemize}
\item Our numerical calculation shows that various types of transition occur even in the presence of magnetic field.
But, opposite to the case of zero magnetic field, in this case the
configuration with the larger value of electric field is
energetically favorable.
\item In the crossover region between NDC and PDC, the value of critical electric field in which non-zero current is observed increases.
\item The type of transition depends on the value of magnetic field as well as the temperature. In fact by increasing the critical magnetic field, the critical temperature also rises.
\item It seems that, at least for the systems we have considered, neither supersymmetry which is preserved by the system at zero temperature nor the dimension of  the subspace on which fundamental matter lives alters the transitions, qualitatively.
\end{itemize}

\section*{Acknowledgment}

We would like to thank D. Allahbakhshi, K. Bitaghsir, A. Davody, H.
Ebrahim and A. E. Mosaffa for useful discussions. We also thank M.
Alishahiha for reading the manuscript and fruitful comments.

The authors thank the referee for his/her positive comments to improve the quality of the paper.


\begin{thebibliography}{99}

\bibitem{Balescu}
R. Balescu, {\em  Equilibrium and Non-equilibrium Statistical
Mechanics} (John Wiley and Sons, 1975)

\bibitem{steady0}
Y. Oono, and M. Paniconi,
{\em Steady state thermodynamics}\/,
Prog. Theor. Phys. Suppl. {\bf 130}, 29--44 (1998).

\bibitem{steady}
  S.~Sasa and H.~Tasaki,
  ``Steady state thermodynamics,''
  [cond-mat/0411052],
 Z.~ Racz,
  ``Nonequilibrium phase transition,''
  [cond-mat/0210435].

\bibitem{steady1}
B.~Derrida,
  ``Non equilibrium steady states: fluctuations and large deviations of the density and of the current,''
  [cond-mat/0703762 ].

\bibitem{Henkel}
M.~Henkel, H.~Hinrichsen and S.~L\"{u}beck, {\sl Non-Equilibrium Phase Transitions. Vol. 1,} (Springer-Verlag, Dordrecht 2008);
M.~Henkel and M.~Pleimling, {\sl Non-Equilibrium Phase Transitions. Vol. 2,} (Springer-Verlag, Dordrecht 2010).

\bibitem{ads/cft}
  J.~M.~Maldacena,
  ``The Large N limit of superconformal field theories and supergravity,''
  Adv.\ Theor.\ Math.\ Phys.\  {\bf 2}, 231 (1998)
  [hep-th/9711200];
  S.~S.~Gubser, I.~R.~Klebanov and A.~M.~Polyakov,
  ``Gauge theory correlators from noncritical string theory,''
  Phys.\ Lett.\ B {\bf 428}, 105 (1998)
  [hep-th/9802109];
  E.~Witten,
  ``Anti-de Sitter space and holography,''
  Adv.\ Theor.\ Math.\ Phys.\  {\bf 2}, 253 (1998)
  [hep-th/9802150];
  J.~McGreevy,
  ``Holographic duality with a view toward many-body physics,''
  Adv.\ High Energy Phys.\  {\bf 2010}, 723105 (2010)
  [arXiv:0909.0518 [hep-th]].

\bibitem{Witten}
  E.~Witten,
  ``Anti-de Sitter space, thermal phase transition, and confinement in gauge theories,''
  Adv.\ Theor.\ Math.\ Phys.\  {\bf 2}, 505 (1998)
  [hep-th/9803131].

\bibitem{Karch}
  A.~Karch and A.~O'Bannon,
  ``Metallic AdS/CFT,''
  JHEP {\bf 0709}, 024 (2007)
  [arXiv:0705.3870 [hep-th]].

\bibitem{O'Bannon}
  A.~O'Bannon,
  ``Hall Conductivity of Flavor Fields from AdS/CFT,''
  Phys.\ Rev.\ D {\bf 76}, 086007 (2007)
  [arXiv:0708.1994 [hep-th]],
  M.~Ali-Akbari and K.~B.~Fadafan,
  ``Conductivity at finite 't Hooft coupling from AdS/CFT,''
  Nucl.\ Phys.\ B {\bf 844}, 397 (2011)
  [arXiv:1008.2430 [hep-th]],

\bibitem{Alam:2012fw}
  M.~S.~Alam, V.~S.~Kaplunovsky and A.~Kundu,
  ``Chiral Symmetry Breaking and External Fields in the Kuperstein-Sonnenschein Model,''
  JHEP {\bf 1204}, 111 (2012)
  [arXiv:1202.3488 [hep-th]];
  N.~Evans, T.~Kalaydzhyan, K.~-y.~Kim and I.~Kirsch,
  ``Non-equilibrium physics at a holographic chiral phase transition,''
  JHEP {\bf 1101}, 050 (2011)
  [arXiv:1011.2519 [hep-th]].

\bibitem{Karch:2002sh}
  A.~Karch and E.~Katz,
  ``Adding flavor to AdS / CFT,''
  JHEP {\bf 0206}, 043 (2002)
  [hep-th/0205236].

\bibitem{Nakamura1}
  S.~Nakamura, ``Negative Differential Resistivity from Holography,''
  Prog.\ Theor.\ Phys.\  {\bf 124}, 1105 (2010)
  [arXiv:1006.4105 [hep-th]].

\bibitem{Nakamura2}
  S.~Nakamura, ``Nonequilibrium Phase Transitions and Nonequilibrium Critical Point from AdS/CFT,''
  Phys.\ Rev.\ Lett.\  {\bf 109}, 120602 (2012)
  [arXiv:1204.1971 [hep-th]].

\bibitem{Kobayashi}
  S.~Kobayashi, D.~Mateos, S.~Matsuura, R.~C.~Myers and R.~M.~Thomson,
  ``Holographic phase transitions at finite baryon density,''
  JHEP {\bf 0702}, 016 (2007)
  [hep-th/0611099].
\bibitem{Mateos}
  D.~Mateos, S.~Matsuura, R.~C.~Myers and R.~M.~Thomson,
  ``Holographic phase transitions at finite chemical potential,''
  JHEP {\bf 0711}, 085 (2007)
  [arXiv:0709.1225 [hep-th]].

\bibitem{Kim:2011qh}
  J.~Erdmenger, R.~Meyer and J.~P.~Shock,
  ``AdS/CFT with flavour in electric and magnetic Kalb-Ramond fields,''
  JHEP {\bf 0712}, 091 (2007)
  [arXiv:0709.1551 [hep-th]];
  C.~Hoyos, T.~Nishioka and A.~Obannon,
  ``A chiral magnetic effect from AdS/CFT with flavor,''
  Lect.\ Notes Phys.\  {\bf 871}, 341 (2013);
  M.~Ali-Akbari and S.~F.~Taghavi,
  ``alpha'-Corrected Chiral Magnetic Effect,''
  Nucl.\ Phys.\ B {\bf 872}, 127 (2013)
  [arXiv:1209.5900 [hep-th]].

\bibitem{Kim:2011qh1}
  K.~-Y.~Kim, J.~P.~Shock and J.~Tarrio,
  ``The open string membrane paradigm with external electromagnetic fields,''
  JHEP {\bf 1106}, 017 (2011)
  [arXiv:1103.4581 [hep-th]].

\bibitem{Erdmenger}
  J.~Erdmenger, N.~Evans, I.~Kirsch and E.~Threlfall,
  ``Mesons in Gauge/Gravity Duals - A Review,''
  Eur.\ Phys.\ J.\ A {\bf 35}, 81 (2008)
  [arXiv:0711.4467 [hep-th]].

\bibitem{CasalderreySolana}
  J.~Casalderrey-Solana, H.~Liu, D.~Mateos, K.~Rajagopal and U.~A.~Wiedemann,
  ``Gauge/String Duality, Hot QCD and Heavy Ion Collisions,''
  arXiv:1101.0618 [hep-th].

\bibitem{Karch:2005ms}
  A.~Karch, A.~O'Bannon and K.~Skenderis,
  ``Holographic renormalization of probe D-branes in AdS/CFT,''
  JHEP {\bf 0604}, 015 (2006)
  [hep-th/0512125].

\end{thebibliography}
\end{document}